# Accelerated Carrier Recombination by Grain Boundary/Edge Defects in MBE Grown Transition Metal Dichalcogenides

*Ke Chen[1], Anupam Roy[2], Amritesh Rai[2], Hema C P Movva[2], Xianghai Meng[1], Feng He[1,3], Sanjay Banerjee[2], and Yaguo Wang[1,3*]*

1. Department of Mechanical Engineering, The University of Texas at Austin, Austin, TX 78712, USA
2. Microelectronics Research Center and Department of Electrical and Computer Engineering, The University of Texas at Austin, Austin, TX 78758, USA
3. Texas Materials Institute, The University of Texas at Austin, Austin, TX 78712, USA

*Corresponding Author: yaguo.wang@austin.utexas.edu

Abstract

Defect-carrier interaction in transition metal dichalcogenides (TMDs) play important roles in carrier relaxation dynamics and carrier transport, which determines the performance of electronic devices. With femtosecond laser time-resolved spectroscopy, we investigated the effect of grain boundary/edge defects on the ultrafast dynamics of photoexcited carrier in MBE grown $MoTe_2$ and $MoSe_2$. We found that, comparing with exfoliated samples, carrier recombination rate in MBE grown samples accelerates by about 50 times. We attribute this striking difference to the existence of abundant grain boundary/edge defects in MBE grown samples, which can serve as effective recombination centers for the photoexcited carriers. We also observed coherent acoustic phonons in both exfoliated and MBE grown $MoTe_2$, indicating strong electron-phonon coupling in this materials. Our measured sound velocity agrees well with previously reported result of theoretical calculation. Our findings provide useful reference for the fundamental parameters: carrier lifetime and sound velocity, reveal the undiscovered carrier recombination effect of grain boundary/edge defects, both of which will facilitate the defect engineering in TMD materials for high speed opto-electronics.

Transition metal dichalcogenides (TMDs), a family of layered materials, have attracted tremendous interest in recent years due to their unique properties at two-dimensional scale, such as direct band gap in monolayer[1], stable exciton[2], strong spin-valley coupling[3], and immunity to short channel effects[4]. One potential application of TMDs is nano- and flexible optoelectronics, in which the dynamics of the photoexcited carriers/excitons plays an essential role in determining the device performance and functionality. Thus, measuring and understanding the dynamics of photoexcited carriers in TMDs is very important to realize novel electronic devices. Currently, there are mainly three ways to produce 2D TMDs: Mechanical exfoliation, Chemical vapor deposition (CVD), and Molecular beam epitaxy (MBE) growth. While the most abundant defects in all three kinds of TMDs are chalcogen vacancies, CVD and MBE TMDs samples show considerable amount of other defect species, such as grain boundary/edge (GB/E) sites and impurities[5]. Typically, defects can have significant impacts on the carrier dynamics. For example, chalcogen vacancies can induce mid-gap states which can give rise to radiative bound excitons, reducing the intrinsic PL intensity[6]; While oxygen impurities occupying chalcogen vacancy sites can eliminate the mid-gap states[7], oxygen impurities taking up molybdenum vacancy sites keep those mid-gap states and play the role as effective carrier trappers[8, 9].

Comparing with exfoliated and CVD synthesized samples, studies on the photoexcited carrier dynamics in MBE grown TMDs are rare. Our previous work has shown that one of the main structural defects in MBE grown TMDs are the GB/E sites[10]. These GB/E defects can hinder carrier transport by introducing localized charge-carrier states. However, the effect of these defects on the carrier relaxation dynamics remains unknown. In this letter, we utilize femtosecond laser pump-probe spectroscopy to investigate the photoexcited carrier dynamics in MBE grown $MoSe_2$ and $MoTe_2$ thin films. Comparing with exfoliated samples, the excited carrier decay rates in MBE grown samples are about 50 times faster, as revealed by transient reflection signals, which suggests that the GB/E defects in MBE samples can serve as effective recombination centers for the photoexcited electron-hole pairs. Coherent acoustic phonons in both MBE grown and exfoliated $MoTe_2$ thin film have also been observed, from which the phonon sound velocities are extracted that agree with the theoretical value.

Hexagonal $MoSe_2$ and $MoTe_2$ thin films (5 nm) are grown by MBE on sapphire substrate. Details of the growth process and the structural, spectroscopic, and electrical characterization of the samples can be found in our previous work[10]. Figure 1a and 1b show the transmission electron microscopy (TEM) images of the MBE samples[10]. The main feature of the MBE samples is the presence of a large amount of nanometer-size grains, and thus, plenty of the GB/E sites. The origin of these small size grains is still not well known, one possible reason could be that the large vapor pressure difference between Mo and the chalcogen species adversely narrows the growth window, thereby making the film prone to chalcogen defects and reduced grain dimensions. Similar grain structures have been observed in MBE MoSe$_2$ reported by other groups[11, 12]. In contrast, exfoliated TMD samples have much larger grain size, and thus much fewer GB/E defects[13-16]. As shown in Figure 1c[14], no obvious GB/E can be seen in exfoliated MoSe$_2$ within an area comparable to Figure 1b.

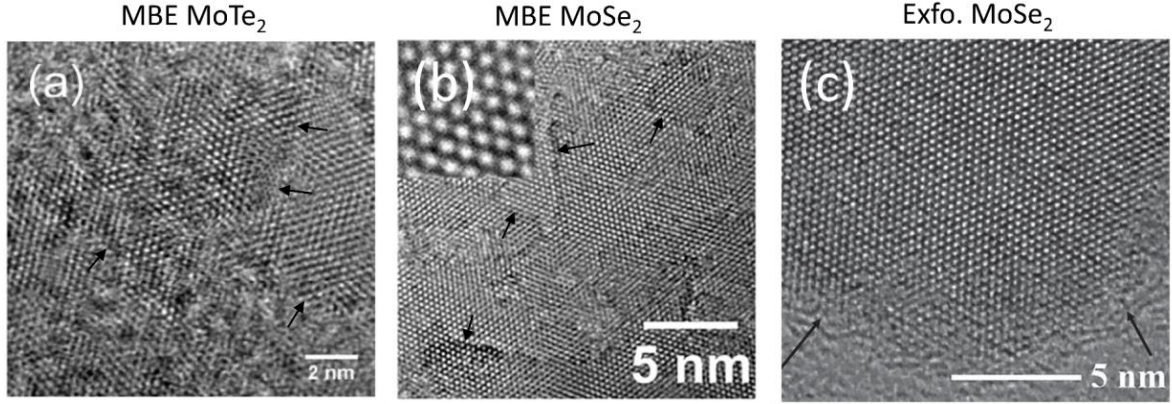

Figure 1. (a) and (b) TEM images of MBE grown MoTe$_2$ and MoSe$_2$[10]. Abundent nano-size grains and plenty of grain boundary/edge defects are present. The arrows indicate some of the grain boundaries and edges. Reprinted with permission from [10]. Copyright © 2016 American Chemical Society. (c) TEM images of exfoliated MoSe$_2$[14]. Very few grain boundary/edge defects can be seen. The arrows indicate the border of the exfoliated sheet. Reprinted with permission from [14]. Copyright 2016 John Wiley & Sons, Inc.

The large difference in the density of GB/E sites between MBE grown and exfoliated samples provides us a good physical model to study the effect of boundary defects on the photoexcited carrier dynamics. For comparison, we have also mechanically exfoliated MoSe$_2$ and MoTe$_2$ thick flakes onto Si and sapphire substrates with scotch tape, respectively. We've measured the transient differential reflection signals ($\Delta R/R_0$, where R and $R_0$ are the reflectivities after and before the pump excitation, respectively) in both MBE grown and exfoliated samples with femtosecond laser pump-probe spectroscopy and compared the carrier relaxation rates in these samples. Our laser pulses are generated from a Ti: Sapphire oscillator operating at 80 MHz repetition rate, with about 100fs pulse width (FWHM), 800 nm central wavelength, and 30 μm diameter of laser spots on the sample surface. Details of our experimental setup can be found in Ref. 8. Figure 2a shows the $\Delta R/R_0$ signals in exfoliated MoTe$_2$ measured at different pump fluences. Besides the sharp rise at the beginning due to the excitation, the signals consist of another small rise after the excitation and then a slow decay component. Noticeable pulse-like signals are observed at the beginning and around 750ps later (echo), which are related to coherent acoustic phonons (strain pulse) generated by pump laser and will be discussed later. Multilayer MoTe$_2$ has an energy difference of 1.16eV at the K point in the momentum space[17]. In our experiment, the pump and probe photon energies are identical (around 1.55eV), much higher than the energy gap at the K point. Therefore, after excitation, the photoexcited electrons/holes will not stay at the pumped (also probed) energy level but quickly relax to the bottom/top states of conduction/valence band through carrier-carrier scattering (carrier thermalization process to reach a temperature defined Fermi distribution) and carrier-phonon scattering (carrier cooling process to reach an equilibrium temperature between carrier and phonon systems), respectively[18]. During these processes, the excess energy of carriers is transferred to the phonon system, generating coherent acoustic phonons and increasing the lattice temperature. After the ultrafast (fs to ps time scale) carrier thermalization[18], phase space filling effect[19] at the probed energy level should be rather weak due to a very small or even no occupation at that level. Hence, unlike the resonantly-probed case, the reflection change ΔR in the non-resonantly probed case here should not be sensitive to the absorption change (imaginary part of complex refractive index) induced by Pauli blocking of the phase filling effect, but mainly dominated by the real part of complex reflective index change induced by the thermalized carriers instead[9]. Therefore, the small increase of $\Delta R/R_0$ signal after the sharp rise can be attributed to the refractive index change (real part) induced by a slight lattice temperature increase, and the decay of $\Delta R/R_0$ signal actually reflects the recombination process of the thermalized electron-hole pairs. Figure 2a shows that the peak of $\Delta R/R_0$ signal is proportional to the pump fluence, and the signals measured at different pump fluences overlap when normalized (as shown in the inset), which indicates that the pump fluences used in our experiments are small enough that the $\Delta R/R_0$ value is linear with the excited carrier density[20].

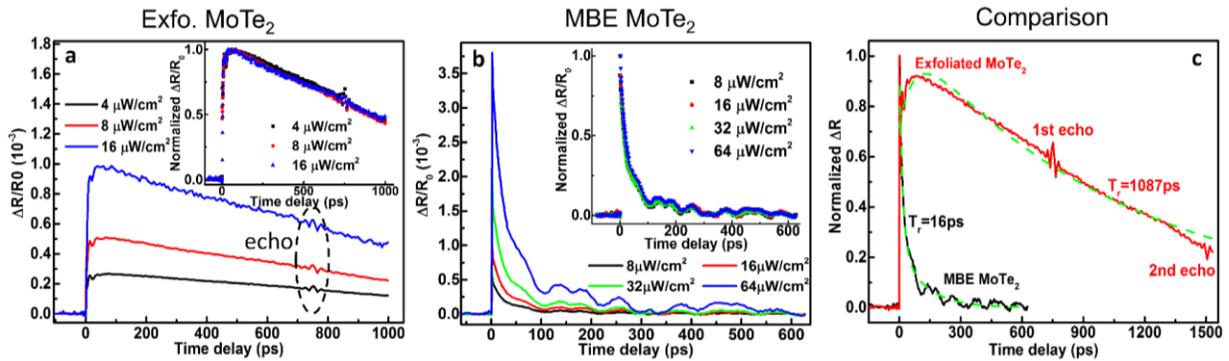

Figure 2. (a) Differential reflection signals of exfoliated MoTe$_2$ at different pump fluences. Inset: normalized $\Delta R/R_0$ signals; (b) Differential reflection signals $\Delta R/R_0$ of MBE grown MoTe$_2$ at different pump fluences. Inset: normalized $\Delta R/R_0$ signals; (c) Comparison of normalized $\Delta R/R_0$ signals in MBE MoTe$_2$ and exfoliated MoTe$_2$, measured at small step size with the echoes and oscillations from acoustic phonons clearly shown. Dashed curves are the fittings using Eq. (1) in the text.

Figure 2b shows the transient differential $\Delta R/R_0$ signals of MBE MoTe$_2$ measured at different pump fluences. The signals are exponential decays superposed with clear periodic oscillations, which are the coherent acoustic phonons (strain pulse) generated by the pump laser and will be discussed later. Similar to that in the exfoliated sample, the decay is from the decreasing of the excited carrier density due to electron-hole pair recombination. The magnitude of $\Delta R/R_0$ signals is also proportional to the pump fluence, and the normalized signals overlap as shown in the inset, indicating that the pump fluence is low enough to ensure a linear relation between the value of $\Delta R/R_0$ signals and the excited carrier density plus the phonon vibration amplitude[20].

The comparison of $\Delta R/R_0$ signals in exfoliated and MBE MoTe$_2$ is shown in Figure 2c. A gigantic difference in the signal decay rates is observed. Based on the above signal analysis, the non-oscillating part of the transient $\Delta R/R_0$ signals (not considering the echoes and oscillations from acoustic phonons) in exfoliated and MBE MoTe$_2$ after carrier thermalization (fs to ps process) can be described with the following expression:

$$\frac{\Delta R}{R_0}(t) = A_T\left(1 - e^{-\frac{t}{\tau_{rise}}}\right)e^{-\frac{t}{\tau_c}} + A_{cool}e^{-\frac{t}{\tau_{cool}}} + A_e e^{-\frac{t}{\tau_r}} \qquad \text{Eq. (1)}$$

The first, the second and the third term on the right-hand side of Eq. (1) represents the lattice temperature change, the carrier cooling process, and the carrier recombination, respectively. $A_T$, $A_{cool}$, and $A_e$ are the signal amplitude for lattice temperature change, carrier cooling, and carrier recombination, respectively, $\tau_{rise}$, $\tau_c$, $\tau_{cool}$, and $\tau_r$ are the time for lattice temperature rise from laser heating, the time for lattice temperature decrease due to heat conduction, the carrier cooling time, and the carrier recombination lifetime, respectively. The fitting results using Eq. (1) are shown in Figure 2c as dashed lines. The fitted value of carrier lifetime $\tau_r$ for exfoliated and MBE MoTe$_2$ are 1087 ps and 16 ps, respectively. Obviously, the huge difference in carrier lifetime should come from the distinction in structural properties between the two samples: either the dimension (thickness) or the structural defects. The MBE sample has a thickness of 5 nm (7 atomic layers). It has been reported that multilayer MoS$_2$ with thickness larger than 5 atomic layers have carrier lifetimes close to that of the bulk MoS$_2$[21], which suggests that the difference in structural defects but not the thickness should be responsible for the observed lifetime difference here. Generally, chalcogen vacancy is the most abundant defect species in TMD materials[22]. However, we can rule out the relevance of Te vacancies to the observed reduction of carrier lifetime in the MBE sample, because the densities of Te vacancies are similar in the exfoliated MoTe$_2$ (Te:Mo ratio ~ 1.90:1, see supplementary information) and the MBE MoTe$_2$ (Te:Mo ratio ~ 1.91:1[10]), based on our X-ray photoelectron spectroscopy (XPS) measurements. As shown in Figure 1, MBE TMDs has much more GB/E defects than the exfoliated one. This distinction strongly suggests that the carrier-GB/E defect interaction should account for the dramatic decrease of carrier lifetime in the MBE MoTe$_2$ sample. GB/E sites are defects in form of displacement of host atoms and represents structural damage to the lattice.

These defects can typically introduce energy levels in the bandgap and serve as efficient recombination centers that can capture the photoexcited electrons and holes at similar rates, with timescale usually much shorter than the intrinsic lifetime[23]. Therefore, the carrier recombination process can be accelerated by the GB/E defects and the effective carrier lifetime is thus shortened dramatically, as demonstrated in Figure 2c. Previously, the defect assisted electron-hole recombination has been observed[24]. However, the previous study could not identify specific type of defect as recombination center. Here, through comparing the carrier relaxation dynamics in exfoliated and MBE grown samples with large difference in the density of specific defects, we suggest that the GB/E defects are responsible for the observed lifetime reduction in MBE grown $MoTe_2$.

To confirm whether the recombination-center nature of GB/E defects is universal in MBE grown TMDs, we conducted similar measurements of the differential reflection signal $\Delta R/R_0$ in exfoliated and MBE grown $MoSe_2$, as shown in Figure 3a and 3b, respectively. Bulk $MoSe_2$ has an energy gap around 1.55 eV at the K(K') point[10], resonant with our pump and probe photon energy. With resonant detection, the reflection change is dominated by the absorption change due to the phase-space filling effect[8, 9]. For both samples (the exfoliated $MoSe_2$ on Si substrate and the $MoSe_2$ thin film grown on sapphire substrate by MBE), the reflection change is proportional to the absorption change, according to the calculation with transfer matrix method[8, 9]. Thus, the response of negative reflection change $\Delta R$ means that the absorption decreases after the pump excitation, which is consistent with Pauli blocking/repulsion induced by phase-space filling effect. As shown in Figure 3a, Figure 3b and the insets, the $\Delta R/R_0$ signals are proportional to pump fluence and overlap when normalized (at around 4 ps), suggesting that the experiments are performed in the linear region and the $\Delta R/R_0$ values is proportional to the excited carrier density at the detected energy states[19]. The sharp peaks shown in Figure 3a and 3b are due to the carrier redistribution after excitation (carrier thermalization and carrier cooling), and the following decay mainly reflects the recombination of carriers in the K(K') valleys[9]. Unlike $MoTe_2$, neither the echoes/oscillations from acoustic phonons, nor significant contribution from lattice heating and heat conduction, is observed in $MoSe_2$. This difference indicates that $MoTe_2$ has strong electron-phonon coupling through which the energy initially absorbed by carrier system can rapidly transfer to phonon system, generate coherent phonons and lead to a prominent lattice temperature increase. It remains unclear why $MoTe_2$ has such an extraordinary electron-phonon coupling strength, but the discussion about this topic is beyond the scope of this article. With weak electron-phonon coupling considered, the $\Delta R/R_0$ signals in $MoSe_2$ samples can also be described by Eq. (1) but without the first (thermal) term. Figure 3c shows the comparison of the normalized $\Delta R/R_0$ signal in the exfoliated and MBE grown $MoSe_2$ samples, along with the fittings with Eq. (1). Similar to $MoTe_2$, the signals in MBE grown $MoSe_2$ also decay much faster than that in the exfoliated sample. The fitted carrier lifetimes are 47 ps and 2206 ps for MBE and exfoliated samples, respectively. This huge difference is also attributed to the acceleration of carrier recombination in MBE grown MoSe$_2$ through the GB/E defects. Thus, the observation of substantially reduced lifetime in MBE grown MoSe$_2$ further confirms the role of GB/E defects as efficient recombination accelerators and this role should be universal in all TMD materials.

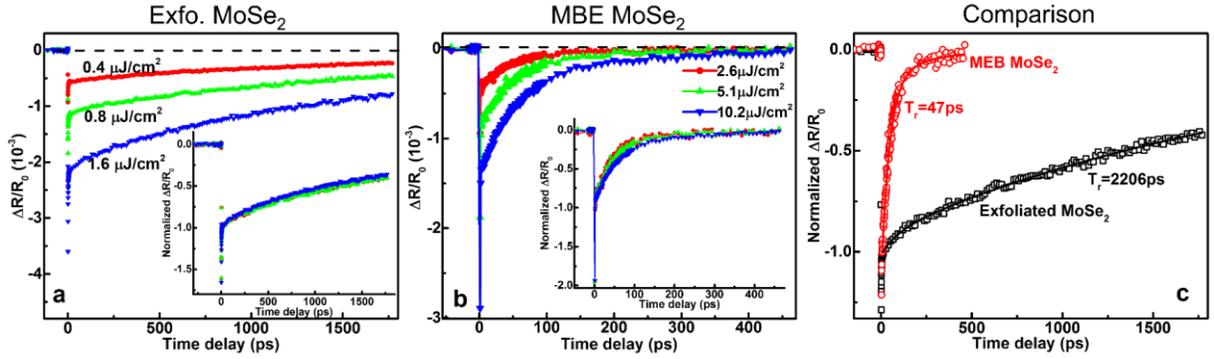

Figure 3. (a) Differential reflection signals of exfoliated MoSe$_2$ samples at different pump fluences. Inset: normalized ΔR/R$_0$ signals; (b) Differential reflection signals ΔR/R$_0$ of MBE MoSe$_2$ samples at different pump fluences. Inset: normalized ΔR/R$_0$ signals; (c) Comparison of normalized ΔR/R$_0$ signals between MBE MoSe$_2$ and exfoliated MoSe$_2$. Dashed curves are the fitting using Eq. (1) in the text.

The oscillations and echoes observed in ΔR/R$_0$ signals reveal the acoustic phonon properties in MoTe$_2$. Figure 4a and Figure 4b show the echo signal in the exfoliated sample and the oscillation signal in the MBE sample, respectively, which are obtained by subtracting the electronic and thermal parts (fitted with Equation 1) from the signals shown in Figure 2c. In the exfoliated sample (Figure 4a), the initially generated coherent phonon (strain pulse) propagates deep into the sample, is reflected at the MoTe$_2$/sapphire interface, and bounce back to the surface as an echo re-detected by the probe light[25]. This process will repeat until the echo amplitude is too weak to detect. The arrival time of the 2$^{nd}$ echo is just about twice of the arrival time of the 1$^{st}$ echo. The thickness of the exfoliated MoTe$_2$ flake $d_{exfo-MoTe_2}$ is around 1.3 μm, which is measured with atomic force microscopy (AFM) (supplementary information). Thus, the sound velocity in exfoliated MoTe$_2$ is estimated as $v_{s_{exfo-MoTe_2}} = 2d_{exfo-MoTe_2}/T_{1st-arrive} = $ 3467 m/s. The oscillations observed in the MBE sample (Figure 4b) also come from the modulation of the refractive index by coherent acoustic phonons. The thickness of MBE sample is only 5 nm, much shorter than the coherence length of the strain pulse (~350 nm, estimated from the pulse duration and the acoustic phonon velocity). The strain wave can bounce back and forth between the MoTe$_2$ surface and the MoTe$_2$/sapphire interface, and form standing waves of coherent acoustic phonons, which can modulate the refractive index and thus the reflection of the probe laser. We can fit the phonon oscillations with two damping cosine functions: $\Delta R_{vib} = A_1 e^{-\frac{t}{\tau_1}}\cos(\omega_1 t + \varphi_1) + A_2 e^{-\frac{t}{\tau_2}}\cos(\omega_2 t + \varphi_2)$, where $A_i, \omega_i, \tau_i$ and $\varphi_i$, are the amplitude, frequency, scattering time and initial phase of the detected phonon modes $i$, respectively. The two fitted frequencies are $\omega_1$=0.0254 THz and $\omega_2$=0.0596 THz. The reason that two phonon frequencies are detected originates from the spectral response of the optical detection of phonon modes. Due to momentum conservation, optical reflectivity detection of coherent phonons is only sensitive to the phonon modes with wave vector $q = 2k_{probe}$, where $k_{probe}$ is the wave vector of the probe laser[26]. Even though the pump laser can generate a bunch of coherent phonon modes (there is also a spectral response for phonon generation), only the phonon modes satisfying $q = 2k_{probe}$ are scattered and detected most efficiently in our reflection signal. Along the cross plane direction, MoTe$_2$ has two acoustic phonon dispersion curves, i.e. the longitudinal acoustic (LA) and transverse acoustic (TA) branches. Therefore, $\omega_1$=0.0254 THz and $\omega_2$=0.0596 THz are probably the TA and LA modes at $q = 2k_{probe} = 2\frac{2\pi}{\lambda_{probe}}$ =1.57 ×10$^7$ 1/m, respectively. Acoustic phonon group velocity is defined as: $v_p = \frac{d\omega}{dq} \approx \frac{\omega}{q}$ (for small $q$). Thus, we can estimate the sound velocity in MBE MoTe$_2$ along the cross-plane direction from the measured frequency $\omega$ and its wave vector $q$ of the LA mode: $v_{S_{MBE-MoTe_2}} = \frac{0.0596 \text{ THz}}{1.57 \times 10^7 \text{ 1/m}} = 3796$ m/s. Our measured cross-plane sound velocities in exfoliated and MBE MoTe$_2$ agree well. To our best knowledge, this is the first time that the cross-plane sound velocity in MoTe$_2$ is experimentally measured by an optical technique. With first-principles calculations[27], sound velocity along the out-of-plane direction in MoTe$_2$ is predicted to be around 2800 m/s . Considering the uncertainty of the DFT calculation on the choice of exchange-correlation functional[27], our experimental sound velocity and the theoretical value agree with each other relatively well.

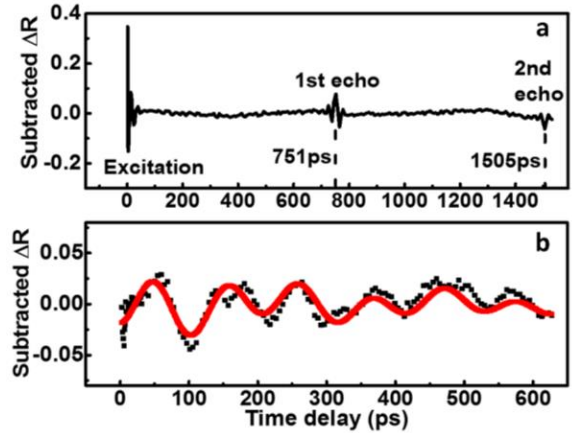

Figure 4. (a) ΔR signal of exfoliated MoTe$_2$ with electron-background subtracted. The 1$^{st}$ and 2$^{nd}$ strain pulse echoes can be observed clearly. (b) ΔR signal of MBE grown MoTe$_2$ with electron-background subtracted. The oscillating feature of the signal comes from the coherent acoustic phonon vibrations that modulate the refractive index of MBE thin film. The red curve is the fitting with two damping cosine functions.

In summary, we have investigated the photoexcited carrier dynamics in exfoliated and MBE grown MoTe$_2$ and MoSe$_2$ with ultrafast laser pump-probe spectroscopy. The comparison of the differential reflection signals between the exfoliated sample and the MBE grown thin films reveals an increase of carrier recombination rate in MBE grown samples by about 50 times. This drastic difference is attributed to the existence of abundant GB/E defects in MBE grown samples. Sound velocities along the cross-plane direction in MoTe$_2$ have been extracted from the coherent acoustic phonon signals, which agree with the theoretical calculation. Our results reveal the acceleration effect of GB/E defects on carrier recombination, provide valuable reference about carrier lifetime in exfoliated and MBE TMDs. The findings here have important implications for defect engineering in TMD materials, especially for applications where short carrier lifetime is critical in determining the device response rate/working frequency, such as TMD-based fast photodetectors, fast saturable absorbers, and bipolar junction transistors working at the switching mode.

Acknowledgement

The authors would like to acknowledge support from National Science Foundation (NASCENT, Grant No. EEC-1160494; CAREER, Grant No. CBET-1351881); Department of Energy (SBIR/STTR, Grant No. DE-SC0013178); and DOD_ Army (Grant No. W911NF-16-1-0559 and MURI Grant No. W911NF-17-1-0312). The authors appreciate technical support from Omicron Company. The authors thank Dr. Tiger H Tao and Shaoqing Zhang for the AFM measurement.

# Supplementary Information

# Accelerated Carrier Recombination by Grain Boundary/Edge Defects in MBE Grown Transition Metal Dichalcogenides

*Ke Chen[1], Anupam Roy[2], Amritesh Rai[2], Hema C P Movva[2], Xianghai Meng[1], Feng He[1,3], Sanjay Banerjee[2], and Yaguo Wang[1,3]**

4. Department of Mechanical Engineering, The University of Texas at Austin, Austin, TX 78712, USA
5. Microelectronics Research Center and Department of Electrical and Computer Engineering, The University of Texas at Austin, Austin, TX 78758, USA
6. Texas Materials Institute, The University of Texas at Austin, Austin, TX 78712, USA

*Corresponding Author: yaguo.wang@austin.utexas.edu

## 1. X-ray photoelectron spectroscopy (XPS) of exfoliated MoTe$_2$ flakes

The elemental composition and chemical stoichiometry were investigated by Omicron Multiprobe X-ray photoelectron spectroscopy (XPS) setup. All XPS spectra were acquired at room temperature using monochromatic Al–Kα (hv = 1486.7 eV) X-ray radiation source[1]. The background pressure during measurements was below $3 \times 10^{-10}$ mbar. The XPS spectra of the exfoliated MoTe$_2$ flake is shown in Figure s1.

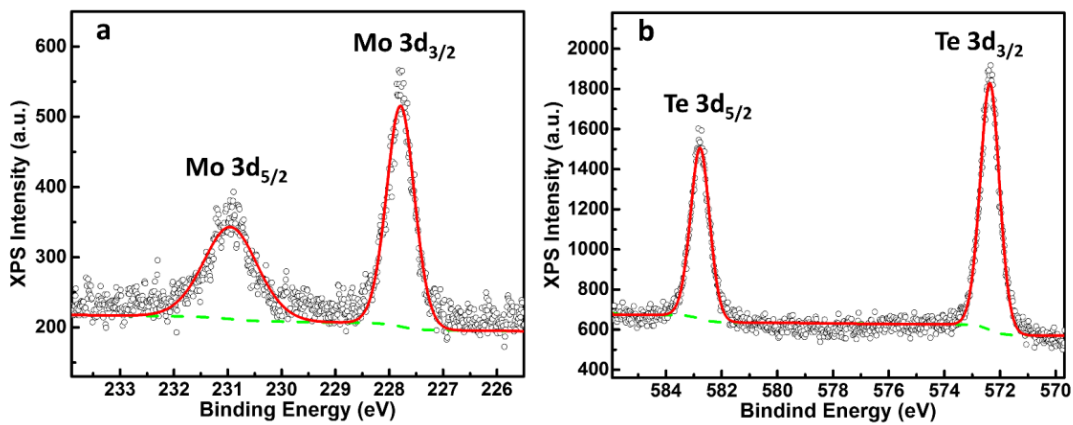

Figure s1. XPS spectra of our exfoliated MoTe$_2$ flakes. (a) shows the Mo 3d peaks while (b) shows the Te 3d peaks. A Te:Mo ratio of 1.90:1 is extracted from the area fit (solid lines) to the experimental data (circles).

## 2. Thickness of the exfoliated MoTe$_2$

The thickness of our exfoliated MoTe$_2$ flake is determined by Atomic force microscope (AFM). The results are shown in Figure s2. The height difference along bar 1 shown in figure s2 indicates that the thickness of the flake is about 1.3 μm.

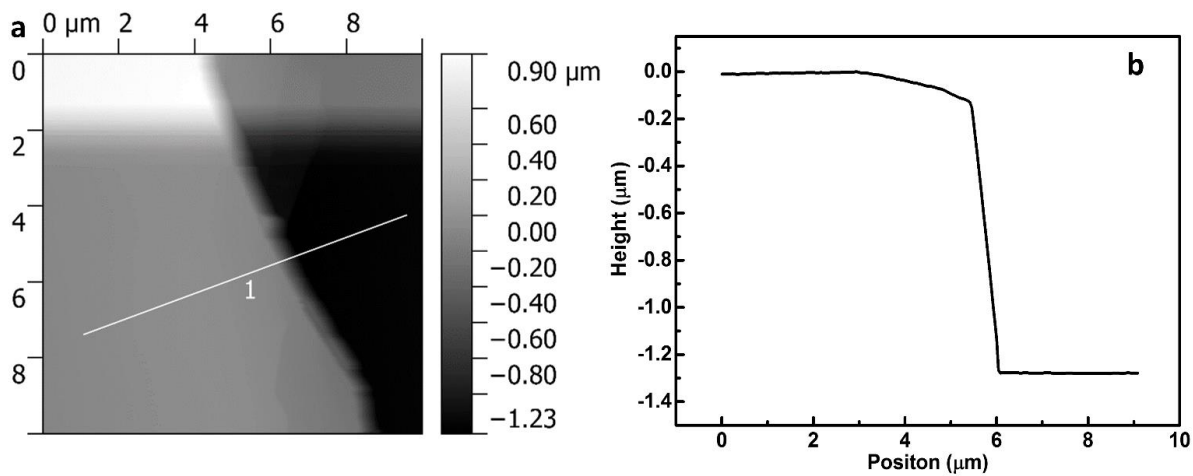

Figure s2. (a) Microscopic image of the exfoliated MoTe2; (b) The height value along bar 1 shown in (a).